\documentstyle[epsf,epsfig,aps]{revtex}
\begin{document}
\preprint{\vbox{\hbox{DOE/ER/40762-187}\hbox{UMD PP\#00-003}}}
\title{Quantum Coins, Dice and Children:\\ Probability and Quantum Statistics}
\author{Chi-Keung Chow and Thomas D.~Cohen}
\address{Department of Physics, University of~Maryland, College~Park, 
MD~20742-4111}
\date{\today}
\maketitle
\begin{abstract}
We discuss counterintuitive aspects of probabilities for systems of identical 
particles obeying quantum statistics.  
Quantum coins and children (two level systems) and quantum dice (many level 
systems) are used as examples.  
It is emphasized that, even in the absence of interactions, 
(anti)symmetrizations of multi-particle wavefunctions destroy statistical 
independences and often lead to dramatic departures from our intuitive 
expectations.  
\end{abstract}

One of the most fundamental differences between classical and quantum 
mechanics is the necessity of introducing quantum statistics --- 
Bose--Einstein statistics \cite{B,E} for particles with integer spins and 
Fermi--Dirac statistics \cite{F,D} for particles with half-integer spins --- 
for systems of identical particles.  
As a part of the standard physics curriculum, quantum statistics are 
usually introduced in courses on statistical physics or quantum mechanics 
with special emphasis on their applications on statistical systems. 
For example, Bose--Einstein statistics provide a natural understanding of 
the blackbody spectrum, and the Fermi theory on electronic bands of condensed 
matter systems has its foundation on Fermi--Dirac statistics.  

On the other hand, probability theories with quantum statistics are rarely 
discussed.  
This is a curious omission, given the close connection between classical 
statistics and probability theory.  
In this letter, we fill this gap by studying several simple examples on 
quantum systems of identical particles with incomplete information.  
The results are often intriguing and counterintuitive.  
It turns out that, by introducing quantum statistics, statistical independence 
between measurements on different particles is lost even in the absence of 
interactions between these particles.  
These examples can be useful in teaching quantum statistics as they 
highlight the differences between classical and quantum statistics.  

\section{Quantum coin tossing}

We will start with the simplest possible example --- the quantum coin tossing 
problem.  
(Our quantum coin tossing problem has little to do with another problem with 
the same name in quantum information theory.)  
Each quantum coin is a particle in one of the two possible quantum states, 
labeled ``heads'' (H) or ``tails'' (T), which are {\it a priori\/} equally 
likely.  
It is clear that the probability of getting a ``heads'' is 50\%, regardless of 
the statistics of the coin.  
Now consider tossing a set of two coins, in that we mean preparing a mixed 
state for which all distinct allowable quantum two-particle states are 
{\it a priori\/} equally likely.  
These conditions are physically realizable for systems with two low-lying 
single particle discrete levels, which are well separated from the other 
levels.  
More specifically, both the energy splitting between the two states, $\delta$, 
and the interaction energy between particles in these states, $\epsilon$, are 
much smaller than the temperature, so that by equipartition both states are 
equally likely to be occupied.   
The temperature is in turn much smaller than $\Delta$, the energy splitting 
between these two low-lying states, and the rest of the spectrum, so that 
these higher states are essentially empty.  
In other words, the temperature $T$ should be chosen in such a way that 
$\epsilon, \delta \ll T \ll \Delta$.  
If such conditions are satisfied, what is the probability the outcome is two 
``heads''?  
The answer depends on which statistics the coins obey.  

$\bullet$  With classical statistics, {\it i.e.}, where the particles are 
distinguishable, there are four possible outcomes: 
\begin{equation}
\hbox{HH}, \qquad \hbox{HT}, \qquad \hbox{TH}, \qquad \hbox{TT}.  
\end{equation}
Since all four outcomes are {\it a priori\/} equally likely, the probability 
for HH is $1/4$.  
This is applicable to tossing macroscopic coins, where quantum effects are 
negligible.  

$\bullet$  With Bose--Einstein statistics, where the allowable states must be 
symmetric under exchange, there are only three possible outcomes: 
\begin{equation}
\hbox{HH}, \qquad (\hbox{HT+TH})/\sqrt{2}, \qquad \hbox{TT}.  
\end{equation}
Consequently, the probability for HH increases to $1/3$.  
This is applicable, for example, to a simple system of two bosons in an 
external potential with doubly degenerate ground states labeled as H and T.  
It is also applicable to two photons in a rectangular optical cavity with 
dimensions $a \times a \times b$ ($a\gg b$).  
Such a cavity has two degenerate ground states, which can be labeled as H and 
T, respectively.  
Then the probability of finding both photons in the H state is $1/3$.  
(This example has been studied in Dirac's {\sl ``The Principle of Quantum 
Mechanics''} \cite{Db}.)

$\bullet$  With Fermi--Dirac statistics the outcomes of HH and TT are 
forbidden as the allowable states must be antisymmetric under exchange; 
there is only one possible state:  
\begin{equation}
(\hbox{HT$-$TH})/\sqrt{2}
\end{equation}
The probability for HH is obviously zero.  
This is applicable to a system of two fermions in an external potential with 
doubly degenerate ground states.  

This above analysis clearly shows that the outcomes of measurements on the two 
coins are not statistically independent.  
Classically, two systems are usually regarded as statistically independent 
if they do not interact with each other.  
This, however, is not necessarily true for quantum mechanical systems of 
identical particles, where the two-particle wavefunction is entangled unless 
it can be written as the product of two single-particle wavefunctions.  
More precisely, for probability applications where one studies mixed states, 
correlations occur unless the two-particle density matrix can be factored 
into two density matrices, each describing one of the particles. 
Bosonic (fermionic) wavefunctions, however, are obtained {\it via\/} 
symmetrization (antisymmetrization) of independent two-particle 
wavefunctions, and such symmetrizations or antisymmetrizations destroy 
statistical independence.  
It is manifestly clear in the case of fermions: the Pauli exclusion 
principle, decreeing that two identical fermions cannot be in the same state, 
is incompatible with statistical independence.  
The analogous effect for bosons is Bose enhancement, which states that 
bosons are more likely to be found in the same state than statistically 
independent particles.  
This simple example of quantum coin tossing illustrates, in a very 
compelling way, the differences between classical and quantum statistics.  

We mention in passing that one can easily generalize the above analysis to 
the following problem.  
For $n$ dice, each equally likely to be any of $k$ state (one of which is 
called ``$\bullet$''), what is the probability that all of them end up being 
in the ``$\bullet$'' state?  
For distinguishable particles there are $k^n$ distinct possible outcomes, and 
the probability for any one of them is $k^{-n}$. 
For fermions the probability for an ``all $\bullet$ state'' is trivially 0 
(for $n>1$), and for bosons it is easy to show that there are $k+n-1\choose n$ 
distinct possible outcomes.  
Since these outcomes are all equally likely, the probability for the 
``all $\bullet$ state'' is $1/{k+n-1 \choose n}$, which is always larger than 
$k^{-n}$.  
In other words, Bose statistics always increases the chance of finding two 
identical bosons in the same state; Bose enhancement is really an enhancement. 
(It is important to note that the above analysis holds if and only if there 
are exactly $k$ accessible states as stated in the problem.  
The answer will be different if, for example, there are $k$ doublets 
({\it i.e.}, $2k$ accessible states) and one of the doublets is labeled 
``heads''.)  

Lastly, a word of caution: real coins and dice do not behave like 
quantum coins and dice --- they are essentially classical objects.  
Coins and dice are always distinguishable from one another, while the 
discussion above is only applicable to indistinguishable particles.  
Even usual quantum systems such as electrons in an external magnetic field 
do not behave as quantum coins as described above.  
The analysis above is valid only if there are exactly two allowable states, 
while an electron in a magnetic field has two spin states {\it for each 
accessible spatial quantum state}.  
There are even more allowable states for real coins and dice, which are 
distinguished not only by the spatial location but also for physical 
variations.  
As a result, the terminologies like ``quantum coins'' should be taken in 
a metaphorical sense only.  

\section{Conditional probabilities: The Quantum Crib}

Now we will move on to conditional probabilities, which are even more  
intriguing and counterintuitive.  
Consider the following famous problem: 

\smallskip
 
(I)  Two children sleep in a crib.  If one is chosen at random and turns 
out to be a boy, what is the probability that both are boys?

(II)  Two children sleep in a crib.  If at least one of them is a boy, 
what is the probability that both are boys?

\smallskip

\noindent The answer is well known: $1/2$ for the question (I), $1/3$ for 
question (II).  
These answers presume that the two genders are {\it a priori\/} equally 
likely, and also that the children are distinguishable objects and 
their genders are statistically independent.  
(Whether these assumptions are strictly true in the real world is  
beyond the scope of this paper.)    
But what if we assume the children obey quantum statistics instead? 
In order to study this question, we will reformulate the above puzzle in the 
following way to make it applicable to quantum particles:  

\smallskip

Consider two identical particles, each being equally probable of being in one 
of two quantum states: either boy (B) or girl (G) at the same spatial 
position, with all distinct allowable gender combinations being {\it a 
priori\/} equally likely.  
(Here the terminologies ``boy'' and ``girl'' are used in a metaphorical sense 
only --- real children are distinguishable classical objects.  
Recall the discussion at the end of the previous section.)  
We will adopt the following shorthand: ``the particle is a B'' stands for 
``the particle is in state B''.  
Then: 

(I)  One particle is selected in a random manner.  
If it is a B, what is the probability that the other one is also a B?  

(II)  Both particles are measured and at least one of them is a B.  
What is the probability that the other is also a B?  

\smallskip

\noindent Both of these questions can be easily answered by listing the 
elements of the spaces of possible combinations.  
For distinguishable children, the space of possible combinations is 
$\{$BB, BG, GB, GG$\}$.  
Out of the four combinations three of them have at least one B, but among 
them only one is BB, the answer to the question (II) is $1/3$, as 
forecasted above.  
On the other hand, since all four combinations are equally probable, and for 
each outcome both particles are equally likely to be selected, there are 
$4\times 2=8$ equally likely cases: 
\begin{eqnarray}
\hbox{{\underline B}B}, &\qquad \hbox{{\underline B}G}, \qquad 
\hbox{{\underline G}B}, \qquad &\hbox{{\underline G}G}, \nonumber\\
\hbox{B{\underline B}}, &\qquad \hbox{B{\underline G}}, \qquad 
\hbox{G{\underline B}}, \qquad &\hbox{G{\underline G}};  
\end{eqnarray}
where the underlined particle is being selected.  
Since in four of these cases B is selected, and among them only two cases 
the remaining particle is a B, the answer to the question (I) is $2/4=1/2$.  
This answer reflects that the two particles are presumed to be statistically 
independent, and knowledge of one of the two children does not have any 
implication for the other child.  

The situation is dramatically changed if these children obey quantum 
statistics instead.  
It is easy to see that for fermionic children, the BB combination is 
forbidden by Pauli exclusion principle, and hence the answer to both 
questions above is: 0.  
For bosonic children, with the space of possible combination being $\{$BB, 
(BG+GB)$/\sqrt{2}$, GG$\}$, two out of the three combination have at least one 
B, and one of them is BB, so the answer to question (II) is $1/2$, in contrast 
to $1/3$ for the case with distinguishable children.  
The analogy of Eq.~(4) is 
\begin{eqnarray}
\hbox{{\underline B}B}, &\qquad 
(\hbox{{\underline B}G+G{\underline B}})/\sqrt{2}, \qquad &
\hbox{{\underline G}G}, \nonumber\\ 
\hbox{B{\underline B}}, &\qquad 
(\hbox{B{\underline G}+{\underline G}B})/\sqrt{2}, \qquad &
\hbox{G{\underline G}}.   
\end{eqnarray}
In three of these cases B is selected, and since in two of them the remaining 
particle is also B, the answer to question (I) is $2/3$, not $1/2$.  
Again, we see that Bose statistics enhances the probability of finding 
two identical bosons in the same state.  

In the above, we have analyzed the problems by listing all the possible 
combinations.  
This becomes less practical for more complicated problems, and one may 
wonder if it is possible to re-analyze these problems in a way which 
can be generalized to more complex settings.  
Since we are studying mixed states, a natural description is {\it via} 
density matrices.  
Both question (I) and (II) will be re-analyzed in the appendix by using the 
density matrices formalism.  
However, the remainder of this paper (except the appendix) is in fact 
accessible without reference to the density matrices formalism.  

\section{Conditional Probabilities: The Quantum Day Care Center}

We will now move on from the crib to the quantum day care center.  
Consider the following problem: 

\smallskip

Consider $n$ quantum children (where $n\gg 1$) in a day care center, where  
by the equal opportunity laws all distinct allowable gender combinations are
{\it a priori\/} equally likely.  
We will define $R$ as the ratio of quantum boys to the total number of 
children in the day care center.  Then 

(III)  What is the probability distribution of $R$?

(IV)  One child is selected in random, which is found to be a boy.  
What is the probability distribution of $R$ for the remaining children?  

\smallskip

\noindent For distinguishable children obeying classical statistics, 
statistical independence implies that the outcome of the remaining $n-1$ 
children are not affected by the outcome of the first child.  
As a result, the probability distribution of $R$ is a sharply peaked 
Gaussian around $R=1/2$ for both questions (III) and (IV).  
On the other hand, one can study this problem for bosonic children by 
enumeration.  
There are $n+1$ distinct allowable gender combinations: 
\begin{equation}
C_k = \{k \; \hbox{boys}, n-k \; \hbox{girls}\}, \qquad 0\leq k \leq n, 
\end{equation}
with all of these combinations {\it a priori\/} equally likely, {\it i.e.}, 
$P(k) = 1/(n+1)$.  
As a result, the probability for $R=k/n$ is $P(R=k/n) = 1/(n+1)$ 
where $k$ is an integer between 0 and $n$ and hence $0\leq R \leq 1$.  
When $n\to\infty$, this approaches the uniform probability distribution: 
\begin{equation}
f_0(R) \equiv {dP(R)\over dR} = 1, \qquad \langle R \rangle_0 \equiv 
\int R f_0(R) dR = 1/2.  
\label{0}
\end{equation}

Question (IV) asks for the probability distribution of $R$ on the condition 
that the first child selected is a boy.  
After the selection, there are only $n-1$ quantum children remaining in the 
quantum day care center, and hence the number of boys left can be any integer 
between 0 and $n-1$.  
Now the probability is $P(k)=1/(n+1)$ for each gender combination $C_k$, 
which after one boy is selected is left with $k-1$ left in the quantum day 
care center, so by Bayes' formula one has 
\begin{eqnarray}
\tilde P(m)&\equiv&P(m\;\hbox{boys left}|\hbox{first child selected is a boy}) 
\nonumber\\&=&P(m+1\;\hbox{boys before selection}|\hbox{first child selected 
is a boy}) \nonumber\\&=&{P(\hbox{first child selected is a boy}|m+1\;
\hbox{boys before selection}) \cdot P(m+1\;\hbox{boys before selection}) 
\over \sum_{j=0}^n P(\hbox{first child selected is a boy}|j\;\hbox{boys before 
selection}) \cdot P(j\;\hbox{boys before selection}) }\nonumber\\ &=& 
{(m+1)/n \times 1/(n+1) \over \sum_{j=0}^n j/n \times 1/(n+1)} = 
{2(m+1)\over n(n+1)}.  
\label{p}
\end{eqnarray}

(We will give a brief description of Bayes' formula for readers who are not 
familiar with probability theory.  
Let $H_j$ ($j=1,\dots,N$) be $N$ mutually exclusive events, with  
probabilities $P(H_j)$.  
Then $P(H_k|A)$, the conditional probability of a particular $H_k$ upon the 
condition that another event $A$ occurs, is given by the Bayes' formula: 
\begin{equation}
P(H_k|A) = {P(H_k) \cdot P(A|H_k) \over \sum_j P(H_j) \cdot P(A|H_j)}.  
\end{equation}  
Discussions of Bayes formula can be found in most standard textbooks on 
probability theory.  
See, for example, Fraser \cite{Fb} or Roe \cite{Rb}.)  

Returning to Eq.~(\ref{p}), one can easily check that the probabilities of 
different possible outcomes add up to unity.  
\begin{equation}
\sum_{m=0}^{n-1} \tilde P(m)= \sum_{m=0}^{n-1} {2(m+1)\over n(n+1)} = 1.  
\end{equation}
The conditional expectation value of $m$ is 
\begin{equation}
\sum_{m=0}^{n-1} m P(m) = \sum_{m=0}^{n-1} {2m(m+1)\over n(n+1)}
={2\over 3} (n-1).  
\end{equation}
Since there are $n-1$ quantum children remaining in the quantum day care 
center, $R=m/(n-1)$, and the conditional expectation value of $R$ is 2/3, 
{\it i.e.}, we expect two-thirds of the remaining quantum children to be boys, 
having determined that a single child (out of a huge day care center) is male!
A little quantum knowledge goes a long way in this problem.  

As the number of children in the quantum day care center tends to infinity, 
{\it i.e.}, $n\to\infty$, the conditional probability $\tilde P$ 
approaches a linear probability distribution: 
\begin{equation}
f_1(R) \equiv {d\tilde P(R)\over dR} = 2R, 
\qquad \langle R \rangle_1 \equiv \int R f_1(R) dR = 2/3,   
\label{1}
\end{equation}
in agreement with the conditional expectation value obtained above.  

As a last example, we generalize the previous case to the quantum die 
rolling problem: 

\smallskip

(V) A quantum die is a quantum mechanical particle, {\it a priori\/} equally 
likely to be in one of $k$ possible states.  
(Again, the terminology ``quantum die'' is used in a metaphorical sense 
only --- real dice are distinguishable classical objects.)  
Consider tossing a set of $n$ quantum dice (where $n\gg 1$), by which we mean 
preparing a mixed state for which all distinct allowable quantum $n$ particle 
states are {\it a priori\/} equally likely.    
Let's label one of the states ``state 1'' and define $R$ to be the fraction 
of quantum dice being in state 1.  
Then $n'$ coins are selected in random and $N_1$ of them turn out to be 
in state 1, $N_2$ of them in state 2, {\it etc.}, such that 
$N_1+N_2+\dots+N_k=n'\ll n$.  
What is the probability distribution of $R$ for the remaining dice?  

\smallskip

\noindent This is a straightforward generalization of questions (III) and 
(IV), which are recovered by setting $k=2$, $N_2=0$ and $N_1=0$ in question 
(III) or $N_1=1$ in question (IV).  
For distinguishable dice, by statistical independence, the probability 
distribution of $R$ is sharply peaked around $1/k$. 
We will show in the appendix, by using the density matrix formalism, that the 
conditional probability distribution of $R$ for the remaining bosonic dice is 
\begin{equation}
f_{N_1,N_2,\dots,N_k}(R) = R^{\nu_1-1} \cdot (1-R)^{\nu_2+\dots+\nu_k-1} / 
B(\nu_1,\nu_2+\dots+\nu_k) , 
\label{n}
\end{equation}
where $\nu_j=N_j+1$ and $B(x,y)$ is the Beta function, and the conditional 
expectation value of $R$ is 
\begin{equation}
\langle R \rangle_{N_1,N_2,\dots,N_k} \equiv \int R f_{N_1,N_2,\dots,N_k}(R) 
dR = \nu_1/(\nu_1+\nu_2+\dots+\nu_k).
\label{nu}
\end{equation}

\section{Discussions}

We emphasize that the examples above are not merely academic but may be
experimentally realizable and testable.  
For example, a Bose--Einstein condensate of $F$-spin-1 atoms ($F$-spin is 
the total spin of the atom, which is the quantum mechanical sum of the total 
angular momentum of the electron system and the nuclear spin) provides a 
natural realization of a system of quantum dice with $k=3$, where the three 
states correspond to $F_z=1$,0 and $-1$ along some axis $\hat z$.  
All distinct allowable combinations of $F$-spins are {\it a priori\/} equally 
likely as long as the system is isotropic, or alternatively the temperature 
is sufficiently high that the anisotropic term in the Hamiltonian is 
negligible, while at the same time being low enough to support a 
Bose--Einstein condensate.  
If such a scenario is realizable, a randomly extracted atom from the 
condensate is equally likely to be in any of the three spin states.  
If the first atom turns out to be in the $F_z=1$ state, however, Eq.~(\ref{nu})
(with $k=3$ and $(N_1,N_2,N_3)=(1,0,0)$) predicts that half of the remaining 
atoms will also be in the $F_z=1$ state.  

In our discussion, we have referred to the particles as ``coins'' (with states 
heads and tail), ``children'' (with states boy and girl) and ``dice'' (with 
states labeled by dots).  
It must be understood that these terminologies are being used in a merely 
metaphorical sense.  
Real children do not spontaneously fluctuate between boy states and girls 
states.  
Macroscopic coins and dices are always distinguishable from one another, both 
by physical variations and by their locations in space.  
Quantum statistics applies only to particles that are indistinguishable and 
sharing the same physical location.

We have seen that counterintuitive results often arise when one tries to 
study probabilities for systems with identical particles obeying quantum 
statistics.  
Given the simplicity of our examples, one may wonder why they are not 
discussed or even mentioned in most undergraduate textbooks on quantum 
physics or statistical mechanics.  
We have attempted a literature search for similar discussions; as far as we 
know, there is no mention of these topics in most standard textbooks on 
quantum mechanics and/or statistical physics.  
On the other hand, as mentioned before, the quantum coin tossing problem with 
two coins was discussed by Dirac in Ref.~\cite{Db}.  
There are also discussions in Griffiths \cite{Gb} and Stowe \cite{Sb} which 
share the philosophy of this paper; the specific examples being considered, 
however, are different from the ones discussed here.  
In particular, none of these discussions studied conditional probabilities, 
which give the clearest and the most counterintuitive manifestation of the 
differences between classical and quantum statistics.  

Returning to the question of why these issues not being brought up in most 
undergraduate textbooks, the reason, we believe, lies in the observation that 
these particularly counterintuitive results occur only in systems with a 
finite number of accessible levels.  
This condition is rarely met in important physical systems; as a 
result, these subtleties are seldom discussed in most undergraduate textbooks,
which understandably tend to focus on systems with more immediate 
applications.  
However, we believe our examples can highlight the differences between 
classical and quantum statistics, and deserve some discussion in 
undergraduate classrooms.  

The main lesson of this discussion is the lack of statistical independence 
between identical particles in quantum statistics.  
For the outcomes of measurements on two particles to be statistically 
independent, the wavefunctions of the two particles must be disentangled.  
In other words, the density matrix describing the two-particle mixed state 
must be factorizable into two separate density matrices, each describing 
one of the particles.  
For identical particles obeying quantum statistics, however, their density 
matrices are always entangled due to (anti)symmetrizations.  
As a result, the outcomes of measurements on identical particles are 
always correlated, violating statistical independence.  

\bigskip

Support of this research by the U.S.~Department of Energy under grant 
DE-FG02-93ER-40762 is gratefully acknowledged. 

\vfill\eject

\centerline{\bf APPENDIX}

\bigskip\bigskip

In this appendix, we will re-analyze questions (I) -- (IV) in the density 
matrices formalism, which carries the advantages of being a systematic 
procedure and can be easily generalized to systems of arbitrary number of 
particles and accessible states.

Let us start with a single quantum coin, with the density matrix:
\begin{equation}
\rho = \textstyle{1\over2} |H\rangle \langle H| + 
\textstyle{1\over2} |T\rangle \langle T|, 
\end{equation}
and for two quantum coins obeying classical statistics ({\it i.e.}, being 
distinguishable), the two-particle density matrix is
\begin{equation}
\rho_{cl} \equiv \rho \otimes \rho = \textstyle{1\over4} |HH\rangle\langle HH| 
+ \textstyle{1\over4} |HT\rangle \langle HT| + \textstyle{1\over4} |TH\rangle 
\langle TH| + \textstyle{1\over4} |TT\rangle \langle TT|, 
\end{equation}
and the coefficient $1\over4$ of the $|HH\rangle \langle HH|$ gives the 
probability of getting two ``heads'' when a set of two coins are tossed.  
Notice that statistical independence is manifest as the two-particle 
density matrix $\rho_{cl}$ is the product of two single-particle density 
matrices $\rho$.  
On the other hand, for quantum coins obeying bosonic (fermionic) statistics, 
the two-particle density matrix is obtained by $\rho_{cl}$ by 
(anti)symmetrization.  
\begin{eqnarray}
\rho_{BE} \equiv & \lambda_{BE} \; S \; \rho_{cl} \; S =& \textstyle{1\over3} 
|HH\rangle \langle HH| + \textstyle{1\over3} |S\rangle \langle S| 
+ \textstyle{1\over3} |TT\rangle \langle TT| \nonumber\\
\rho_{FD} \equiv & \lambda_{FD} \; A \; \rho_{cl} \; A =& |A\rangle \langle A| 
\; ; 
\end{eqnarray}
where $S$ and $A$ are the symmetrization and antisymmetrization projection 
operators, respectively; the $\lambda$'s are normalization constants to 
ensure that the density matrices are properly normalized, {\it i.e.}, 
${\rm Tr}\rho_{BE} = {\rm Tr}\rho_{FD} =1$, and 
\begin{equation}
|S\rangle = (|HT\rangle + |TH\rangle)/\sqrt{2}, \qquad
|A\rangle = (|HT\rangle - |TH\rangle)/\sqrt{2}.  
\end{equation}
Again, the probabilities of getting two ``heads'' can be read off as the 
coefficients of the operator $|HH\rangle \langle HH|$.  
The coefficients are $1/3$ and 0 for bosonic and fermionic statistics, 
respectively, confirming the values obtained through listing.  
After all, the density matrix formalism is simply a systematic way to generate 
and organize the list of all possible combinations, with the probability 
of each  combination appearing as the coefficient of the respective 
projection operator.  

\bigskip

As for the problems on conditional probabilities, the condition decreed in 
question (II), namely at least one of the quantum children is a boy, can 
be imposed by projecting out the subspace of ``all girls'' by the projection 
operator $P = {\bf 1} - |GG\rangle \langle GG|$, with {\bf 1} denoting the 
identity operator.  
Acting $P$ onto the two-particle density matrices $\rho_{BE}$ and $\rho_{FD}$ 
above (and renaming ``H'' as ``B'' and ``T'' as ``G''), 
the projected density matrices are,  
\begin{eqnarray} 
\bar\rho_{BE} &\equiv \bar\lambda_{BE} \; P \; \rho_{BE} \; P = & 
\textstyle{1\over2} |BB\rangle \langle BB| + 
\textstyle{1\over2} |S\rangle \langle S|, \nonumber\\
\bar\rho_{FD} &\equiv \bar\lambda_{FD} \; P \; \rho_{FD} \; P = & 
|A\rangle \langle A|,  
\end{eqnarray}
where the $\bar\lambda$'s are again normalization constants.  
Again, the conditional probabilities of BB can be easily read off.  

The condition decreed in question (I), that a randomly chosen quantum child 
turns out to be a boy, can be imposed by using the ``boy annihilation 
operator'' $a_B$, satisfying 
\begin{equation}
a_B |\hbox{$m_B$ boys, $m_G$ girls}\rangle = 
\sqrt{m_B} |\hbox{$m_B-1$ boys, $m_G$ girls}\rangle.  
\label{aB}
\end{equation}
After randomly taking a child out of the crib and find that it is a boy, 
the density matrix of the remaining child is 
\begin{eqnarray} 
\tilde\rho_{BE} &\equiv \tilde\lambda_{BE} \; a_B \; \rho_{BE} \; a_B^\dag = & 
\textstyle{2\over3} |B\rangle \langle B| + 
\textstyle{1\over3} |G\rangle \langle G|, \nonumber\\
\tilde\rho_{FD} &\equiv \tilde\lambda_{FD} \; a_B \; \rho_{FD} \; a_B^\dag = & 
|G\rangle \langle G|,  
\end{eqnarray}
where as before $\tilde\lambda$'s are normalization constants.  
The conditional probabilities of the remaining quantum child is a boy (so that 
both quantum children are boys) again appear as coefficients.  

\bigskip

In the remainder of this appendix, we derive the answers (\ref{0}) and 
(\ref{1}) to the quantum day care center problem for bosonic children.  
Instead of tackling questions (III) and (IV) specifically, we will study 
the more general question (V).  
Questions (III) and (IV) are recovered by setting $k=2$, $N_2=0$ and 
$N_1=0$ in question (III) or $N_1=1$ in question (IV).  

Recall that the ``state 1 annihilation operator'', $a_1$, annihilates a 
quantum die in state 1, and one can analogously define $a_j$ for other 
states, with $1\leq j \leq k$.  
For any {\it complex\/} unit vector $\vec r = (r_1,\dots,r_k)$ satisfying 
$\sum_{j=1}^k |r_j|^2=1$, the linear combination $A_{\vec r} = \vec r \cdot 
\vec a$ (where $\vec a = (a_1, \dots, a_k)$) satisfies $[A_{\vec r}, 
A_{\vec r}^\dag] = 1$ (with $\hbar$ also set to unity), and is the 
annihilation operator for a quantum die in a particular state described by the 
``polarization vector'', $\vec r$.  
It is convenient to parameterize the components of $\vec r$ in the following 
way.  
\begin{eqnarray}
r_1 &=& e^{i\alpha_1} \, \cos\theta_1, \nonumber\\
r_2 &=& e^{i\alpha_2} \, \sin\theta_2 \, \cos\theta_2, \nonumber\\
\vdots & & \qquad \vdots \\
r_{k-1} &=& e^{i\alpha_{k-1}} \, \sin\theta_1 \, \sin\theta_2 \dots 
\sin\theta_{k-2} \, \cos\theta_{k-1}, \nonumber\\
r_k &=& e^{i\alpha_k} \, \sin\theta_1 \, \sin\theta_2 \dots 
\sin\theta_{k-2} \, \sin\theta_{k-1}. \nonumber
\end{eqnarray}
Note that, while a {\it real\/} unit vector in ${\bf R}^k$ lies on a 
$k-1$-dimensional sphere and hence is described by $k-1$ angles $\theta_j$, 
a {\it complex\/} unit vector in ${\bf C}^k$ lies on a $2k-1$-dimensional 
sphere and $k$ extra phases $\alpha_j$ are needed.  

The density matrix of a state with $n$ atoms ($n\gg1$), all 
polarized in the $\vec r$ direction, is given by, 
\begin{equation}
\rho_{\vec r} = (1/n!)\; 
(A_{\vec r}^\dag)^n |0\rangle \langle 0| (A_{\vec r})^n.  
\end{equation}
However, as stated in the problem, all distinct allowable combinations 
are equally likely.   
As a result, the density matrix $\rho$ for such a state will be a 
superposition of $\rho_{\vec r}$ for all $\vec r$.  
\begin{equation}
\rho_0={\Gamma(k/2)k!\over2\pi^{3k/2}n!}\;\int dr_1\,dr_1^*\,\dots dr_k\,dr_k^*
\;\delta(|\vec r|-1)\;(A_{\vec r}^\dag)^n |0\rangle \langle 0| (A_{\vec r})^n, 
\end{equation}
where the asterisks represent complex conjugation and $\Gamma(k/2)k!/
(2\pi^{3k/2}n!)$ is an overall normalization factor to ensure that 
${\rm Tr} \,\rho_0=1$. 
Since the angles $\alpha_j$ are the phases of $r_j$, this density matrix can 
be rewritten as 
\begin{equation}
\rho_0={2^k\Gamma(k/2)\over 2\pi^{k/2}}{k!\over n!}\;\int_0^\infty
|r_1|\,d|r_1|\,\dots\,|r_k|\,d|r_k|\;\int_0^{2\pi} {d\alpha_1\over 2\pi}\,
\dots\, {d\alpha_k\over 2\pi}\;\delta(|\vec r|-1)\;(A_{\vec r}^\dag)^n 
|0\rangle \langle 0| (A_{\vec r})^n. 
\end{equation}
Then we can express the real unit vector $(|r_1|,\dots,|r_k|)$ in terms of 
the angles $\theta_j$.  
Integrating over the Dirac delta distribution $\delta(|\vec r|-1)$ gives a 
factor of $2\pi^{k/2}/(2^k\Gamma(k/2))$, and $\rho_0$ can be recast as 
\begin{equation}
\rho_0 = {1\over n!} \; \int d\Omega_{\bf C} \;
\int_0^{2\pi} {d\alpha_1\over 2\pi}\, \dots\,{d\alpha_k\over 2\pi}\;
(A_{\vec r}^\dag)^n|0\rangle\langle 0|(A_{\vec r})^n = \int d\Omega_{\bf C} 
\; \int_0^{2\pi} {d\alpha_1\over 2\pi}\,\dots\, {d\alpha_k\over 2\pi}\; 
\rho_{\vec r}, 
\label{dm}
\end{equation}
where
\begin{equation}
d\Omega_{\bf C}\equiv \prod_{j=1}^{k-1} dP^{(j)} \equiv \prod_{j=1}^{k-1} 
(k-j+1) \sin^{k-j}\theta_j \cos\theta_j \, d\theta_j,\qquad  
\int_{\theta_j=0}^{\theta=\pi/2}  dP^{(j)} = 1. 
\end{equation}
The interpretation of Eq.~(\ref{dm}) is clear.  
The measure $d\Omega_{\bf C}$ gives the probability distribution 
$f^{(j)}(\theta_j)$ in the domain $[0,\pi/2]$.  
\begin{equation}
f^{(j)}(\theta_j) \, d\theta_j \equiv (k-j+1) \sin^{k-j}\theta_j \cos\theta_j, 
\qquad \int_0^{\pi/2} f^{(j)}(\theta_j) d\theta_j=1. 
\end{equation} 
On the other hand, the phases $\alpha_j$ are equally likely to take any value 
between 0 and $2\pi$.  
\begin{equation}
f(\alpha_j) = 1/2\pi.
\end{equation}
Note that the probability distribution of all three angles $\theta_j$ and 
the phases $\alpha_j$ are independent of each other.  

We are interested in evaluating the expectations of the number operators 
$n_1$ under the density matrix (\ref{dm}).  
It is convenient to introduce the observable $R=n_B/n$, denoting the fraction 
of quantum dice in state 1.   
Notice that this definition of $R$ coincides with that in problem (III).   
Since $n_1 = n \cos^2\theta_1$, the observable $R$ can be re-expressed in 
terms of the angles $\theta_1$ as $R=\cos^2\theta_1$.   
It is straightforward to rewrite $dP^{(1)}(\theta_1)$ in terms of $R$. 
\begin{equation}
d\Omega_{\bf C} = f_{0,\dots,0}(R)dR, \qquad 
f_{0,\dots,0}(R)=(1-R)^{k-2}/(k-1), 
\end{equation}
where the subscripts remind us that no die of any state has been removed.  
In particular, for $k=2$, we have the answer to question (III): the 
probability distribution of $R$ is given by $f_0(R)=1$.   

Now, with the distribution function $f_{0,\dots,0}(R)$, it is straightforward 
to solve problem (V), which is to evaluate the conditional probability 
distribution given that $n'$ coins have been removed, and among them $N_j$ 
of them are found to be in state $j$.  
The density matrix after the selection, which can be written as $\rho_{N_1, 
N_2,\dots,N_k} = \lambda \; (a_1^{N_1} a_2^{N_2} \dots a_k^{N_k}) \, \rho_0 \, 
(a_1^{N_1} a_2^{N_2} \dots a_k^{N_k})^\dag$, where $\lambda$ is a 
normalization constant and $\rho_0$ is the density matrix defined in 
Eq.~(\ref{dm}).  
Since $\rho_1 \neq \rho_0$, the {\it conditional\/} probability distribution 
of $R$ is no longer given by $f_{0,\dots,0}(R)$, but instead 
\begin{eqnarray}
f_{N_1,N_2,\dots N_k}(R) &=& {R^{N_1} \cdot (1-R)^{N_2+\dots+N_k} 
f_{0,\dots,0}(R) \over \int_0^1 R^{N_1} \cdot(1-R)^{N_2+\dots+N_k} 
f_{0,\dots,0}(R) dR} \nonumber\\ &=& {R^{\nu_1-1} \cdot 
(1-R)^{\nu_2+\dots+\nu_k-1} \over B(\nu_1,\nu_2+\dots+\nu_k)}, 
\end{eqnarray} 
reproducing Eq.~(\ref{n}) with $\nu_j = N_j +1$.  
In particular, with $k=2$ and $(N_1,N_2) = (1,0)$, question (V) reduces to 
question (IV) with the answer 
\begin{equation}
f_1(R) = 2R, 
\end{equation}
which agrees with Eq.~(\ref{1}).


\begin{thebibliography}{99}
\bibitem{B} S.N.~Bose, {\sl ``Planck's law and light quanta hypothesis,''} 
Z.~Phys.~{\bf 26}, 178---181 (1924). 
\bibitem{E} A.~Einstein, {\sl ``Quantentheorie des einatomigen idealen 
gases,''} Ber.~Akad.~Wiss.~Berlin 261---267 (1924); 3---14, 18---25 (1925).  
\bibitem{F} E.~Fermi, {\sl ``Zur quantelung des idealen einatomigen gases,''} 
Z.~Phys.~{\bf 36}, 902---912 (1926).  
\bibitem{D} P.A.M.~Dirac, {\sl ``On the theory of quantum mechanics,''} 
Proc.~Roy.~Soc.~{\bf A112}, 661---677 (1926).  
\bibitem{Db} P.A.M.~Dirac, {\sl ``The principles of quantum mechanics,''} 
Clarendon Press (Oxford 1967).  
\bibitem{Fb} D.A.S.~Fraser, {\sl ``Probability and statistics: theory and 
applications,''} Duxbury Press, (North Scituate, Massachusetts 1976).  
\bibitem{Rb} B.R.~Roe, {\sl ``Probability and statistics in experimental 
physics,''} Springer-Verlag (New York 1992).  
\bibitem{Gb} D.J.~Griffiths, {\sl ``Introduction to quantum mechanics,''} 
Prentice Hall (Englewood Cliffs, New Jersey 1995).  
\bibitem{Sb} K.~Stowe, {\sl ``Statistical mechanics and thermodynamics,''} 
John Wiley and Sons (New York 1984).  
\end{thebibliography}
\end{document}